\documentclass[12pt, a4paper]{article}
\usepackage{graphicx}
\usepackage{soul}
\usepackage{xcolor}

\usepackage[square,numbers,comma]{natbib}

\title{Essay written for the Gravity Research Foundation 2021 Awards for Essays on Gravitation: Tully - Fisher Relations and Retardation Theory for Galaxies}

\author{\normalsize
Asher Yahalom$^{a,b}$\\
\footnotesize $^a$ Ariel University, Ariel 40700, Israel\\
\footnotesize $^b$ Princeton University, Princeton, New Jersey 08543, USA\\
\footnotesize e-mail: asya@ariel.ac.il
}
\date{\footnotesize \today}

\begin{document}
\maketitle

\newcommand{\beq} {\begin{equation}}
\newcommand{\enq} {\end{equation}}
\newcommand{\ber} {\begin {eqnarray}}
\newcommand{\enr} {\end {eqnarray}}
\newcommand{\eq} {equation}
\newcommand{\eqs} {equations }
\newcommand{\mn}  {{\mu \nu}}
\newcommand{\ab}  {{\alpha \beta}}
\newcommand{\sn}  {{\sigma \nu}}
\newcommand{\rhm}  {{\rho \mu}}
\newcommand{\sr}  {{\sigma \rho}}
\newcommand{\bh}  {{\bar h}}
\newcommand{\br}  {{\bar r}}
\newcommand {\er}[1] {equation (\ref{#1}) }
\newcommand {\ern}[1] {equation (\ref{#1})}
\newcommand {\Ern}[1] {Equation (\ref{#1})}

\begin{abstract}
Galaxies are huge physical systems having dimensions of many tens of thousands of light years. Thus any change at the galactic center will be noticed at the rim only tens of thousands of years later.
Those retardation effects seem to be neglected in present day galactic modelling used to calculate rotational velocities of matter in the rims of the galaxy. The significant differences between the predictions of Newtonian theory and observed velocities are usually explained by either assuming dark matter or by modifying the laws of gravity (MOND). In this essay we will show that taking retardation effects into account one can explain the azimuthal velocities of galactic matter and the well known Tully-Fisher relations of galaxies.
\end{abstract}

\vfill

\subsection* {Introduction}

The Tully–Fisher relation is an empirical relationship between the mass or intrinsic luminosity of a spiral galaxy and its asymptotic rotation velocity or emission line width. It was first published in 1977 by astronomers R. Brent Tully and J. Richard Fisher \cite{TF}. The relation states that baryonic galactic mass is proportional to velocity to the power of roughly four.

Dynamics of large scale structures is inconsistent with Newtonian mechanics. This was notified in the 1930's by Fritz Zwicky \cite{zwicky},
who pointed out that if more (unseen) mass would be present one would be able to solve the apparent contradiction. The phenomena was also observed in galaxies by Volders \cite{volders} who have shown that star trajectories near the rim of galaxies do not move according to Newtonian predictions, and later corroborated by Rubin and Ford~\cite{rubin1,rubin2,Binney} for spiral galaxies.

In a series of papers we have shown that those discrepancies result from retarded gravity as dictated by the theory of general relativity \cite{YaRe1,ge,YaRe2,YahalomSym,Wagman}. Indeed in the absence of temporal density changes, retardation does not effect the gravitational force. However, density is not constant for galaxies, in fact there are many processes that change the mass density in galaxies over time. Mass accretion from the inter galactic medium and internal processes such as super novae leading to super winds \cite{Wagman} modify the density.

Here we show that the Tully–Fisher relation can be deduced from retardation theory.

\subsection* {Linear Approximation of GR}

Except for the extreme cases of compact objects (black holes and neutron stars) and the very early universe (big bang) one need not consider
the full non-linear Einstein equation \cite{YaRe1}. In most other cases of astronomical interest (galactic dynamics included) one can linearize those equations around the flat Lorentz metric $\eta_{\mn}$ such that\footnote{Private communication with the late Professor Donald Lynden-Bell}:
 \beq
 g_{\mn} = \eta_{\mn} + h_{\mn}, \quad \eta_{\mn} \equiv \ {\rm diag } \ (1,-1,-1,-1), \quad |h_{\mn}|\ll 1
 \label{lg}
 \enq
 One than defines the quantity:
 \beq
 \bar h_\mn \equiv h_\mn -  \frac{1}{2} \eta_\mn h, \quad h = \eta^{\mn} h_{\mn},
 \label{bh}
 \enq
 $\bar h_\mn = h_\mn $ for non diagonal terms. For diagonal terms:
 \beq
 \bar h = - h \Rightarrow  h_\mn = \bar h_\mn -  \frac{1}{2} \eta_\mn \bar h .
 \label{bh2}
 \enq
  It can be shown (\cite{Narlikar} page 75 exercise 37, see also \cite{Edd,Weinberg,MTW}), that one can  choose a gauge such that the Einstein equations are:
 \beq
\bh_{\mn, \alpha}{}^{\alpha}=-\frac{16 \pi G}{c^4} T_\mn , \qquad \bh_{\mu \alpha,}{}^{\alpha}=0.
\label{lineq1}
\enq
\Ern{lineq1} can always be integrated to take the form \cite{Jackson}\footnote{For reasons why the symmetry between space and time
is broken see \cite{Yahalom,Yahalomb,Yahalomc,Yahalomd}}:
 \ber
& & \bh_{\mn}(\vec x, t) = -\frac{4 G}{c^4} \int \frac{T_\mn (\vec x', t-\frac{R}{c})}{R} d^3 x',
\nonumber \\
 t &\equiv& \frac{x^0}{c}, \quad \vec x \equiv x^a \quad a,b \in [1,2,3], \quad \vec R \equiv \vec x - \vec x', \quad R= |\vec R |.
\label{bhint}
\enr
The factor before the integral is small: $\frac{4 G}{c^4} \simeq 3.3 \ 10^{-44}$ hence in the above calculation
one can take $T_\mn$ which is zero order in $h_\ab$.
Let us now calculate the affine connection in the linear approximation:
\beq
\Gamma^\alpha_\mn = \frac{1}{2} \eta^\ab \left(h_{\beta \mu, \nu} + h_{\beta \nu, \mu} - h_{\mn, \beta}\right).
\label{affinel}
\enq
The affine connection has only first order terms, hence for a first order approximation of
$\Gamma^\alpha_\mn u^\mu u^\nu$ appearing in the geodesic,  $u^\mu u^\nu$ is zeroth order. In the zeroth order:
\beq
u^0=\frac{1}{\sqrt{1-\frac{v^2}{c^2}}},  \qquad u^a = \vec u =\frac{\frac{\vec v}{c}}{\sqrt{1-\frac{v^2}{c^2}}} , \qquad
\vec v \equiv  \frac{d \vec x}{d t}, \quad v= |\vec v|.
\label{uz}
\enq
For non relativistic velocities:
\beq
u^0 \simeq 1,  \qquad \vec u \simeq \frac{\vec v}{c} , \qquad u^a \ll u^0   \qquad {\rm for} \quad v \ll c.
\label{uzslo}
\enq
Inserting \ern{affinel} and \ern{uzslo} in the geodesic equation we arrive at the approximate form:
\beq
\frac{d v^a}{dt}\simeq - c^2 \Gamma^a_{00} = - c^2 \left( h^a_{0,0} - \frac{1}{2} h_{00,}{}^a \right)
\label{geol}
\enq
Let us now look at $T_\mn = (p+\rho c^2) u_\mu  u_\nu - p \ g_\mn$. In the current case $\rho c^2 \gg p$, combining this with  \ern{uzslo} we arrive at $T_{00} = \rho c^2 $ while all other components of the tensor
$T_\mn$ are significantly smaller. This implies that $\bar h_{00}$ is significantly larger than other components of the tensor
$\bar h_\mn$. Of course one should be careful and not deduce from the different magnitudes of quantities that such a difference
exist between their derivatives. In fact by the gauge condition in \ern{lineq1}:
\beq
\bar h_{\alpha 0,}{}^0=-\bar h_{\alpha a,}{}^a \qquad \Rightarrow
\bar h_{00,}{}^0=-\bar h_{0 a,}{}^a, \quad \bar h_{b0,}{}^0=-\bar h_{b a,}{}^a.
\label{gaugeim}
\enq
Hence the zeroth derivative of $\bar h_{00}$ (contains a $\frac{1}{c}$ factor) is the same order as the spatial derivative
of $\bar h_{0a}$ and like wise the zeroth derivative of $\bar h_{0a}$ (which appears implicitly in \ern{geol}) is the same order
of  the spatial derivative of $\bar h_{ab}$. However, it is safe to compare spatial derivatives of $\bar h_{00}$ and $\bar h_{ab}$
and conclude that the former is significantly larger than the later. Using \ern{bh2} and taking the above consideration into account
we write \ern{geol} as:
\beq
\frac{d v^a}{dt}\simeq \frac{c^2}{4} \bar h_{00,}{}^a \Rightarrow \frac{d \vec v}{dt} = - \vec \nabla \phi = \vec F,
\qquad \phi \equiv \frac{c^2}{4} \bar h_{00}
\label{geol2}
\enq
Thus $\phi$ is a gravitational potential of the motion which can be calculated using \ern{bhint}:
\beq
\phi = \frac{c^2}{4} \bar h_{00}
= -\frac{ G}{c^2} \int \frac{T_{00} (\vec x', t-\frac{R}{c})}{R} d^3 x' = -G \int \frac{\rho (\vec x', t-\frac{R}{c})}{R} d^3 x'
\label{phi}
\enq
and $\vec F$ is the force per unit mass. If $\rho$ is static we are in the realm of the Newtonian instantaneous action at a distance theory.
However, it is unlikely that $\rho$ is static as a galaxy will attract mass from the intergalactic medium.

\subsection* {Beyond the Newtonian Approximation}

\begin{figure}
\centering
\includegraphics[width=0.8\columnwidth]{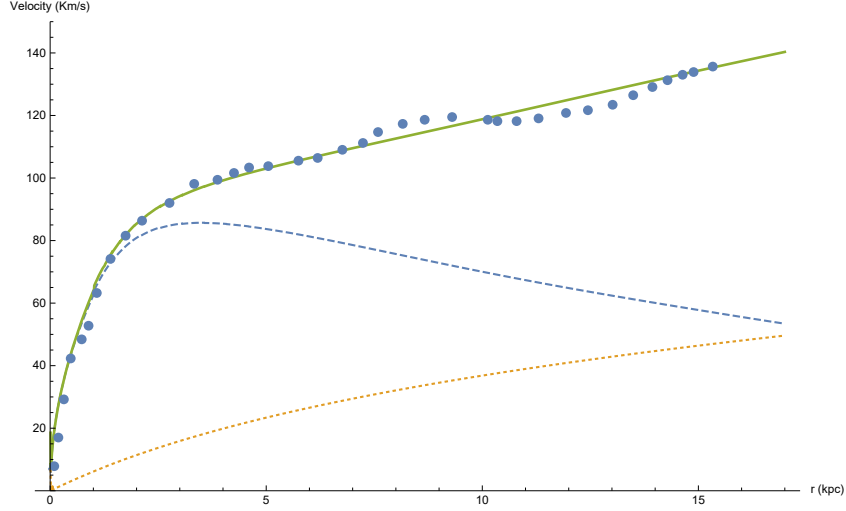}
 \caption{Rotation curve for M33. The~observational points were supplied by Dr. Michal Wagman, a~former PhD student at Ariel University, under~my supervision, using~\cite{Corbelli2}; the full line describes the complete rotation curve, which is the sum of the dotted line, describing the retardation contribution, and~the dashed line, which is the Newtonian~contribution.}
 \label{vcrhoc2}
\end{figure}

The retardation time $\frac{R}{c}$ which may be a few tens of thousands of years is short with respect
to the time that the galactic density changes significantly. This means that we can write a Taylor series for the density:
\beq
\rho (\vec x', t-\frac{R}{c})=\sum_{n=0}^{\infty} \frac{1}{n!} \rho^{(n)} (\vec x', t) (-\frac{R}{c})^n,
\qquad \rho^{(n)}\equiv \frac{\partial^n \rho}{\partial t^n}.
\label{rhotay}
\enq
Inserting \ern{rhotay} into \ern{phi} and keeping the first three terms we will obtain:
\beq
\phi = -G \int \frac{\rho (\vec x', t)}{R} d^3 x' +  \frac{G}{c}\int \rho^{(1)} (\vec x', t) d^3 x'-
\frac{G}{2 c^2}\int R \rho^{(2)} (\vec x', t) d^3 x'
\label{phir}
\enq
The first term will provide the Newtonian potential, the second term does not contribute, the third term will result in the lower order correction to the Newtonian theory:
\beq
 \phi_r = - \frac{G}{2 c^2} \int  R \rho^{(2)} (\vec x', t) d^3 x'
\label{phir2}
\enq
The~expansion given in Equation~(\ref{phir}), being a Taylor series expansion up to the second order, is only valid for limited~radii:
\beq
R < c \ T_{max} \equiv R_{max}
\label{Rmax}
\enq
hence the current approximation can only be used in the near field regime, this is to be contrasted with the far field approximation used for gravitational radiation \cite{Einstein2,Taylor,Castelvecchi}.
The total force per unit mass is:
\ber
\vec F &=& \vec F_N + \vec F_r
\nonumber \\
 \vec F_N &=& - \vec \nabla \phi_N =  -G \int \frac{\rho (\vec x',t)}{R^2} \hat R d^3 x', \qquad \hat R \equiv \frac{\vec R}{R}
\nonumber \\
 \vec F_r &\equiv& - \vec \nabla \phi_r =  \frac{G}{2 c^2} \int  \rho^{(2)} (\vec x', t) \hat R d^3 x'
\label{Fr}
\enr
While the Newtonian force $\vec F_N$ is always attractive the retardation force $\vec F_r$ can be
either attractive or repulsive. Also notice that while the Newtonian force decreases as $\frac{1}{R^2}$ , the retardation force is independent of distance as long as the Taylor approximation of \ern{rhotay} is valid. For short
distances the Newtonian force is dominant but as the distances increase the retardation force becomes dominant. Newtonian force can be
neglected for distances significantly larger than the retardation distance:
\beq
 R \gg R_r \equiv c \Delta t
\label{Rr}
\enq
$\Delta t$ is the typical duration associated with the second temporal derivative of $\rho$. Of course for $R\ll R_r$ the retardation effect can be neglected and only Newtonian forces should be considered. To calculate the  rotation curve we will use equation (23) of \cite{YahalomSym} such that:
\beq
\frac{v_\theta^2}{\br} =  F,
\label{Eulersr3}
\enq
in the above $v_\theta$ is the azimuthal velocity, $\br$ is the cylindrical radial coordinate and $F$ is given in \ern{Fr}. The results for the galaxy M33 \cite{YahalomSym} are depicted in figure \ref{vcrhoc2},
which yields a perfect fit for $R_r = 4.54$ kpc (we remark that the fit does not require tweaking the mass to light ratio, as is done by other authors). Other excellent fits for different types of
galaxies can be found in \cite{Wagman}.

\subsection* {The Tully-Fisher Relations}

\begin{figure}
\centering
\includegraphics[width=0.8\columnwidth]{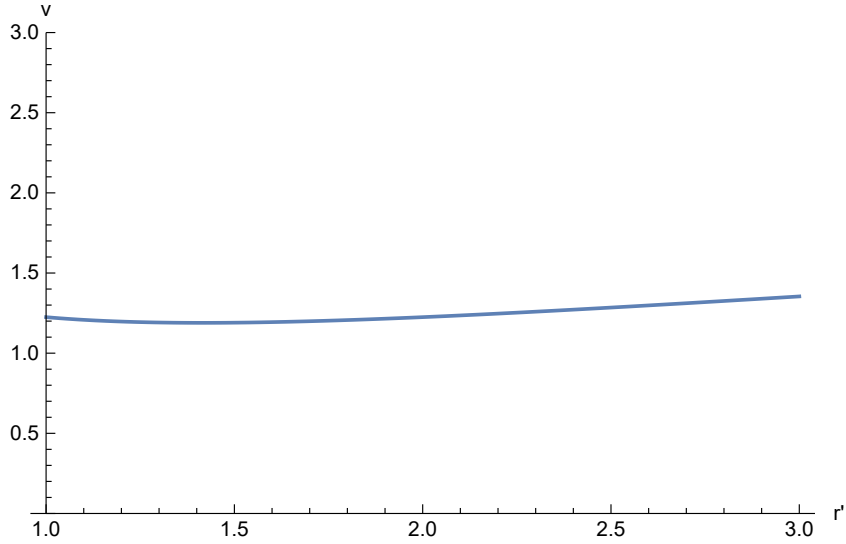}
 \caption{Normalized velocity curve for $r'\in [1,3]$}
 \label{vn}
\end{figure}

For large distances $r=|\vec x|\rightarrow\infty$ such that
 $\hat R \simeq \frac{\vec x}{|\vec x|} \equiv \hat r$ we obtain:
\beq
\vec F_r =  \frac{G}{2 c^2} \hat r \int  \rho^{(2)} (\vec x', t)  d^3 x' =  \frac{G}{2 c^2} \hat r \ddot{M}, \qquad
\ddot{M} \equiv \frac{d^2 M}{dt^2}.
\label{Fr2}
\enq
And:
\beq
\vec F_N  =  -G \int \frac{\rho (\vec x',t)}{R^2} \hat R d^3 x' = -G \frac{M}{r^2} \hat r
\label{FN2}
\enq
in which $M$ is the galactic mass. Now as the galaxy attracts intergalactic gas its mass increases thus $\dot{M}>0$,
however, as the intergalactic gas is depleted the rate at which the mass increases must decrease hence $\ddot{M}<0$. Thus in the
galactic case:
\beq
\vec F_r =  - \frac{G}{2 c^2}  |\ddot{M}| \hat r
\label{Fr3}
\enq
and the retardation force is attractive. The asymptotic form of the total gravitational force is:
\beq
\vec F = \vec F_N + \vec F_r  = - \frac{G M}{r^2}  \hat r \left (1 + \frac{|\ddot{M}|}{2 M c^2} r^2 \right) =  - \frac{G M}{r^2}  \hat r \left (1 + \frac{r^2}{2 R_r^2} \right)
\label{Ft3}
\enq
in which $R_r \equiv c \sqrt{\frac{M}{|\ddot{M}|}}$ (see equation (35) in \cite{YahalomSym}). Assuming
the calculation is done in the galactic plane, we may write the squared azimuthal velocity using \ern{Eulersr3} as:
\beq
v_\theta^2 =   \frac{G M}{r}  \left (1 + \frac{r^2}{2 R_r^2} \right).
\label{Eulersr4}
\enq
Introducing the typical velocity $v_{ty} \equiv \sqrt{\frac{G M}{R_r}}$ and the dimensionless quantities: $v \equiv \frac{v_\theta}{v_{ty}},  r' \equiv \frac{r}{R_r}$ we obtain:
\beq
v^2 =   \frac{1}{r'}  \left (1 + \frac{r'^2}{2} \right).
\label{Eulersr5}
\enq
This form is of course only valid provided the second order approximation holds which is limited
by $R_{max}$ according to \ern{Rmax}. This form is also invalid within the galaxy itself as we assume $r=|\vec x| \gg |\vec x'| $, a detailed calculation of the radial velocity within the galaxy itself (and nearby) is given in \cite{YahalomSym}. Thus the range of validity of the above expression is rather limited. Taking into account the above caveats we assume that most of the galactic baryonic mass is inside  $R_r$, we shall also assume that $R_{max} \approx 3 R_r$, thus the above expression can be taken as a {\bf rough} approximation for the velocity curve in the range $r'\in [1,3]$ as depicted in figure \ref{vn}.
Notice that the plot is almost flat, however, it has a shallow minimum at $r'_{min}=\sqrt{2}$ for which $v_{min} \approx 1.19$. The maximal value in this range is obtained for $r'_{max}=3$ and is
$14 \%$ higher. Thus we can assume that roughly $v \approx 1$ and hence:
\beq
v_{\theta} \approx  v_{ty} = \sqrt{\frac{G M}{R_r}} = \sqrt{\frac{G M}{c \sqrt{\frac{M}{|\ddot{M}|}}}} = \sqrt{\frac{G }{c}}M^{\frac{1}{4}} |\ddot{M}|^{\frac{1}{4}}
\label{vth}
\enq
From which we deduce the Tully-Fisher relation:
\beq
M = k v_{\theta}^4, \qquad k \approx \frac{c^2}{G^2 |\ddot{M}|}
\label{tulf}
\enq
The proportionality constant $k$ will depend on the specific type of the galaxy and its unique circumstances through $|\ddot{M}|$ as is well known (see also \cite{Wagman}).

\section*{Acknowledgment}

The author would like to thank his forme student Dr. Michal Wagman for bringing the issue
of the Tully-Fisher relation to his attention.

\end{document}